# Origin and photoluminescence of a shallow hydrogen-related donor in silicon

Miaomiao Han,[1,2*] Peter Deák,[2,3] Bing Huang,[3] and Adam Gali[2,3,4*]

[1] School of Science, Huzhou Normal University, Huzhou, 313000 Zhejiang, China
[2] HUN-REN Wigner Research Center for Physics, Budapest, PoB 49, H-1525 Budapest, Hungary
[3] Beijing Computational Science Research Center, 100193 Beijing, China
[4] Department of Atomic Physics, Budapest University of Technology and Economics, H-1521 Budapest, Hungary

Corresponding authors: mmhan923@mail.ustc.edu.cn, gali.adam@wigner.hun-ren.hu

ABSTRACT

Paramagnetic color centers in silicon are being investigated recently for use in quantum technologies. While the origin of most of the prominent photoluminescence (PL) lines are well understood, that is not the case for those connected to hydrogen, even though, being a versatile tool in silicon technology, it is almost always present in silicon. E.g., shallow hydrogen-related donors (SHDs) are being used to create deep, buried junctions but their atomic structure remains elusive. Recent experiments suggested connection between the SHDs and the so-called X and the $B_i$ (or $X_i$) PL-lines. An SHD, termed the MT2 center, was tentatively assigned to a bistable, negative-U defect, arising due to the interaction of hydrogen with the self-interstitial complex $I_4$. Using high-level density functional theory calculations, we investigate the interaction of hydrogen with self-interstitial complexes $I_n$. A shallow effective-mass-theory donor arises when hydrogen attacks a Si–Si bond in $I_n$, inserting itself as a bond-center interstitial ($I_nH_{BC}$) in an exothermic reaction. We demonstrate this in the case of a single self-interstitial and of a complex of four. The latter $I_4H_{BC}$ complex shows electronic, spin and PL properties remarkably close to MT2. We find that hydrogen in the presence of $I_4$ follows the same behavior as in pristine silicon, i.e., it is a negative-U bistable defect, only detectable under illumination. In the $I_4H_{BC}$ complex, H can hop via photo-assisted mechanism between equivalent BC sites of the $I_4$ complex, leading to a motional averaged tetragonal symmetry as experimentally observed for MT2. The zero-phonon line is distinct from the X-line (attributed to the bare $I_4$ complex), and falls well into the range of the $X_i$ lines. Our results identify the $I_4H$ complex as a prototype for a class of SHDs, inferring that similar SHDs may arise, with the $X_i$ lines as their PL fingerprints, when H attacks $I_n$ clusters.

## INTRODUCTION

Due to the unprecedented fabrication capabilities for silicon-based electronics and photonics, defects of the silicon crystal are receiving again increasing attention lately from the viewpoint of quantum information technology.[1,2,3,4,5] Recent work revealed that isolated, single color centers in Si could serve as on-demand sources of single telecom photons[6] and as light-matter interface,[7,8] fundamental building blocks for advanced quantum technologies including quantum communication and computation.[9] While the origin of most of the prominent color centers in silicon, usually labeled as G-, C-, and W-lines, are already known by now [G-center: ($C_{Si}$-I-$C_{Si}$),[2,6,10,11,12,13,14,15,16,17,18,19] W-center: $I_3$,[20] and C-center: ($C_i$+$O_i$),[21,22,23] where I stands for a self-interstitial and *i* for an interstitial impurity], that is not the case for those connected to hydrogen containing defects, except for the T-center [$(C_2)_{Si}$+H].[7,24,25,26,27,28] The so called X- and $X_i$ (sometimes called $B_i$) lines in the PL spectrum all have been attributed at one point or another to hydrogen complexes but the structure of those have never been identified (we hereafter refer collectively to this unified class of similar defects as $X_i$ lines).[29,30,31,32]

Apart from the possibility of future use in quantum information technology, hydrogen plays already an important role in the actual technology of electronic devices. Among others, it can produce shallow hydrogen-related donors (SHDs), routinely used for creating junctions far from the surface.[33,34] SHDs arise in proton-irradiated or in radiation-damaged and hydrogen-treated silicon, after annealing at temperatures between ~300 and ~500 °C, while they disappear above 550°C.[35,36,37,38] SHDs have been observed, independent of doping, in both Czochralksy (Cz-) and floating zone (FZ- ) Si, even in high-resistivity, high-purity samples. Investigations during the last quarter of the 20th century implicated that there are several defect types grouped together under the term SHD. A clear distinction among the different types is made difficult, on the one hand, by the fact that structural (e.g., electron paramagnetic resonance, EPR) or dynamical (annealing) studies have hardly ever been coupled with spectroscopy, [e.g., deep level transient spectroscopy (DLTS), photoluminescence (PL), or infrared (IR) absorption], in the same investigation, and usually only the total free carrier concentration was monitored either by electrical or by resonance methods. On the other hand, during the heat treatment, which is necessary to produce the SHDs, the radiation-induced defects themselves, to which hydrogen atoms can couple, do evolve or disappear. A PTIS (photothermal ionization spectroscopy) study, in which hydrogenation was carried out by applying hydrogen-plasma treatment to un-annealed irradiated silicon, has revealed seven SHDs with vertical ionization energies between 34 and 53 meV.[39] Except for SHDs related to the bistable $C_iO_i$ complex,[40] the atomistic origin for the rest could not been unambiguously elucidated so far. Another type of SHDs is usually attributed to the interaction of H with intrinsic defects, more specifically with self-interstitial complexes, but their atomistic structure is not known either.[41,42] Here we investigate such defects and establish their electronic and photoluminescence (PL) properties.

Earlier we have shown that the X-line can unambiguously be assigned to a complex of four self-interstitials ($I_4$).[43] Recently, the temperature evolution of the SHD concentration and the PL spectrum was studied simultaneously.[29] It has been shown that SHDs anneal out in two stages, first between 380-440 °C and then between 500-540 °C. The decay in the second stage suggested correlation with the X-center, so the SHDs with higher thermal stability were interpreted as hydrogen(s) associated with the $I_4$ defect. However, new PL lines were also observed, with the ones at 1124, 1130 and 1137 meV becoming prominent at or above 460 °C, i.e., in the second stage. Note that, those lines are close to the $X_i$ lines observed between 1137-1150 meV earlier.[31,32] , Kioyi and Umeda then reported electron paramagnetic resonance (EPR) experiments on SHDs in proton-irradiated, P-doped (830 Ωcm) FZ-Silicon, identifying three EPR signals with $S$=1/2.[30] The one termed MT1 could be observed in the dark, MT2 and MT3 (with higher thermal stability) could only be observed under photoexcitation. The latter two were attributed to the same defect, with MT3 arising when two close-by MT2 defects interact. The g-tensors showed tetragonal symmetry, and the $^{29}$Si hyperfine (hf) interaction of MT2 was similar to the EPR center called $B_3$, which is usually considered to originate from the tetra-interstitial cluster $I_4$ (see Ref.[43] and references therein). Therefore, the authors assigned these EPR centers to the $I_4$ defect interacting with different number of captured hydrogen atoms, even though the hf interaction with hydrogen nuclei couldn't be observed. Free carrier concentrations have been deduced from spreading resistance measurements, and since the measured total spin concentration was about 80-85% of the free carrier concentration, it was claimed that the MT signals arise from SHDs. MT1 was not observed after heat treatment at 500 °C, so it was assumed that it belongs to SHDs which anneal out in the first stage (380-440 °C), while MT2 to SHDs annealing out in the second stage (500-540 °C). The cyclotron resonance (CR) spectrum taken at 15K showed the presence of free carriers in both samples (annealed at 400 or 500 °C). Since MT1 was not detected in the 500°C sample, it was assumed that at 15K the free carriers in both samples are provided by MT2 only. Since MT2 was not detected at lower temperatures, it was assumed, that it is a negative-U double donor, i.e., its donor state is either doubly occupied or empty, and the paramagnetic, singly-occupied state can only be produced by illumination. The MT1 signal was detectable even in the dark, however, the intensity diminished somewhat below 15K. Therefore it was assumed that the ground state of MT1 is also S=0. Apparently, MT1 is related to deeper SHDs which anneal out in the lower temperature range and, according to Figs.7-8 of Ref.[29], in correlation with the C-line of the PL spectrum. It has been shown that the C-line originates from the $C_iO_i$ complex.[21] Since the $C_iO_iH$-type SHDs also anneal out at ~460 °C,[40] it seems reasonable to assume that MT1 belongs to one of those.

By means of high-level density functional theory calculations, we investigate here the interaction of hydrogen with self-interstitial complexes $I_n$, therefore, MT2 is of more interest to this paper. However, the assignment of MT2 to a double-donor $I_4H_2$ complex has some problems. First, the tetragonal symmetry of $I_4$ could only be maintained in $I_4H_2$ if both H atoms were located on the main axis of $I_4$, but in that case a $H_2$ molecule would form with no EPR signal. Second, if MT2 was a double donor then the free carrier concentration should be about twice the sum of total spin concentration. We note here that the EPR signal of the bond-centered hydrogen interstitial (AA9) was also detected only under illumination. This was explained by electron transfer between $H_{BC}$ and some radiation defect: either donation, resulting in a proton at the BC site, or capture, creating a negatively charged interstitial hydrogen, which would then move to the AB (or T) site.[44] DLTS studies have unambiguously shown the latter to be the case.[45] As we will show, a similar interpretation can be applied in the case of the MT2 center as well, without assuming it to be a double donor. We find that the hydrogen atom attacks a Si–Si bond in $I_n$, inserting itself as a bond-center defect ($I_nH_{BC}$) giving rise to a shallow donor state. This model is verified in $IH_2H_{BC}$ and $I_4H_{BC}$. The calculated photoluminescence of the latter is close to the $X_i$ lines and to the 1124~1137 meV lines of Ref.[29]. The calculated hyperfine tensor also fits that of the MT2 center well.

## COMPUTATIONAL METHODS

Calculations have been carried out using the Heyd-Scuseria-Ernzerhof (HSE06) functional[46] as implemented in the Vienna *ab-initio* simulation package VASP.[47,48] The defect was modelled in a 512 atom silicon supercell with a plane wave cutoff of 420 eV, and all atoms were allowed to relax in a constant volume till the forces were below 0.01 eV. The Γ-point sampling of the Brillouin zone was used. For the description of core electrons, projector augmented wave pseudopotentials were used.[49] To get a better description of H-related defects, a Hubbard $U_s = -8.5$ V on-site potential in the Dudarev approach[50,51] was applied to the *s* orbital of the H atom, as described in detail in the Supplemental Material (SM). Excited states were calculated with the ΔSCF (or cDFT) method. To get the correct PL energy, a spin correction in the excited state was applied.[52] In case of the bound exciton, a band-filling correction was also used,[53] since the Γ-point in the reduced Brillouin-zone of the supercell represents the [0.75,0,0] primitive k-point explicitly, while the real CBM is at [0.83,0,0]. Charge correction was also used in this case, since the electron wave function is delocalized, similarly to the jellium counter charge to a positive defect. Furthermore, the singularity in the Coulomb-kernel for exact-exchange is circumvented in VASP by an expansion with auxiliary functions,[54] but using the $\Gamma$-point alone causes an artificial deepening of conduction-band-derived occupied states.[55] Therefore, the total energy of the excited state was corrected with the energy difference of this level and its unoccupied counterpart.[56] Hyperfine parameters were calculated with the VASP implementation of Marsman.[57] Diffusion energy barriers were obtained by the climbing image nudged elastic band (CI-NEB) method.[58] For that, as well as for calculating local vibration modes, the computationally affordable Perdew-Burke-Ernzerhof (PBE) functional[59] was used. For the formation energy of defects, the following equation were used:

$$E^q_{form} = E^q - E_{host} - \sum_i n_i \mu_i + q(E_V + E_F) + E^q_{corr} \quad [1]$$

where $E^q$ is the calculated total energy of the supercell with a defect in the charge state $q$, and $E_{host}$ without the defect. $n_i$ is the number of atoms of type $i$, introduced into the supercell when forming the defect, and $\mu_i$ is the chemical potential of that element. For H that was set to be half of the total energy of the $H_2$ molecule (at 0K), while for Si we used $E_{host}/512$. $E_F$ is the Fermi energy, $E_V$ is the valence band maximum (VBM) in the bulk. $E^q_{corr}$ is the energy correction of the charged defect obtained by applying the SCPC method.[60]

RESULTS

As it is well known (and reproduced in the SM), a hydrogen atom in the pristine silicon lattice prefers to sit into the bond between two Si atoms (bond center, or BC site) in its positive charge state, and behind a bond (antibonding, or AB site) when it is negative.[61] The neutral state at the BC site is metastable. The charge transition level (CTL) at the BC site, E(+/0) = $E_{CBM}$ – 0.18 eV, is relatively deep. It is to be expected, however, that hydrogen at the BC site in one of the compressed Si–Si bonds of a self-interstitial complex ($I_n$) will be shallower. We demonstrate this in the case of a single self-interstitial. In the pristine lattice, the self-interstitial (I) is a [110]-oriented pair of Si atoms sharing a lattice site, $I_{[110]}$. This is a highly mobile defect[62] but interaction with hydrogen turns it into a tilted $I_{[001]}H_2$ configuration and immobilizes it.[63,64,65] Our results show that inserting a proton between the two Si atoms, $I_{[001]}H_2H_{BC}^+$ (as shown in Fig. S1), is highly exothermic with respect to isolated $I_{[001]}H_2$ and $H_{BC}^+$. The $I_{[001]}H_2H_{BC}^0$ defect produces an EMT (effective mass theory) shallow donor state indeed, with a calculated vertical CTL of 0.06 eV. To our knowledge, such an $I_{[001]}H_2H_{BC}^0$ defect has not been identified so far, but – just as $H_{BC}$ – it will probably be detectable in EPR only under illumination. In Table S3 (in the SM) we provide the calculated hyperfine tensor.

Aggregates of self-interstitials ($I_n$) also typically contain such tilted [001] pairs,[66] so similar behavior can be expected when hydrogen attacks an $I_n$ complex. We investigate this in the case of $I_4$ in detail, due to the availability of experimental data. $I_4$ contains four nearby split self-interstitials, slightly tilted from [001], which form a cage (see Fig.1).[67-69] Exploring possible sites of H both in the neutral and in the positive charge states, we found that H in the BC site of a horizontal Si–Si bond (as shown in Fig. 1a) is energetically most favorable. The formation of this $I_4H_{BC}$ complex from isolated $H_{BC}(0)$ and $I_4$, will release 0.22 eV, so H would easily get trapped at $I_4$. In the negative charge state, H prefers an AB site next to the $I_4$ complex, but outside the cage (as shown in Fig. 1b) , which is 0.05 eV lower in energy than inside.

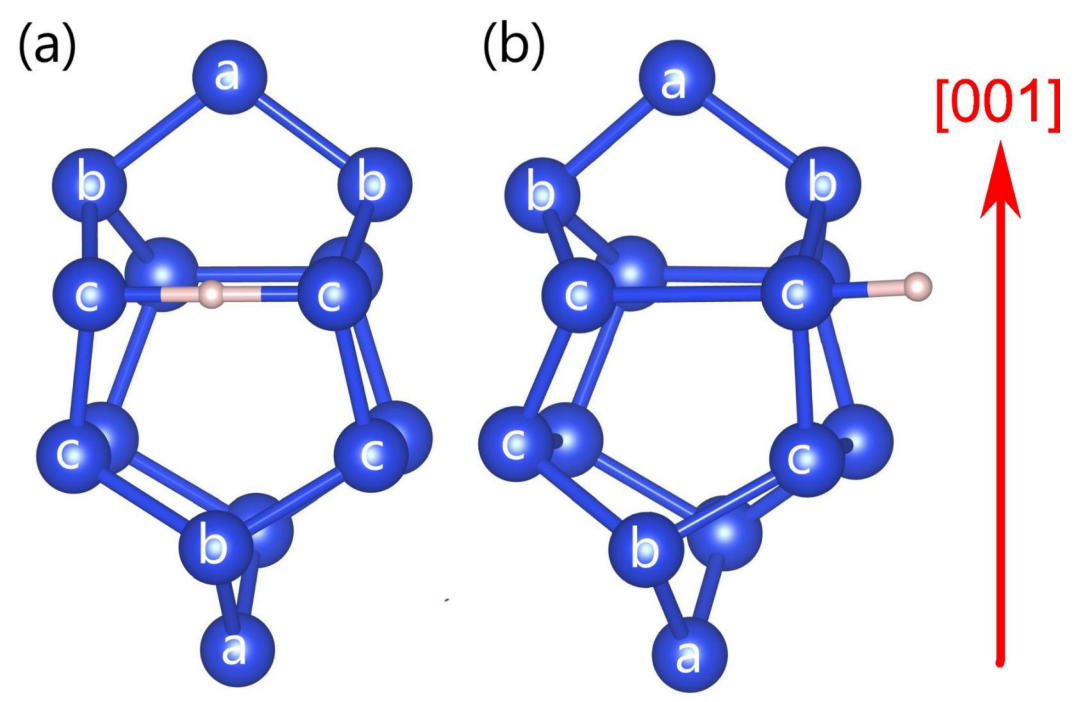

Fig. 1. Geometry of H(+) at the BC site of $I_4$ (a), and of H(−) at the AB site next to $I_4$ (b). The red arrow indicates the [001] crystallographic direction.

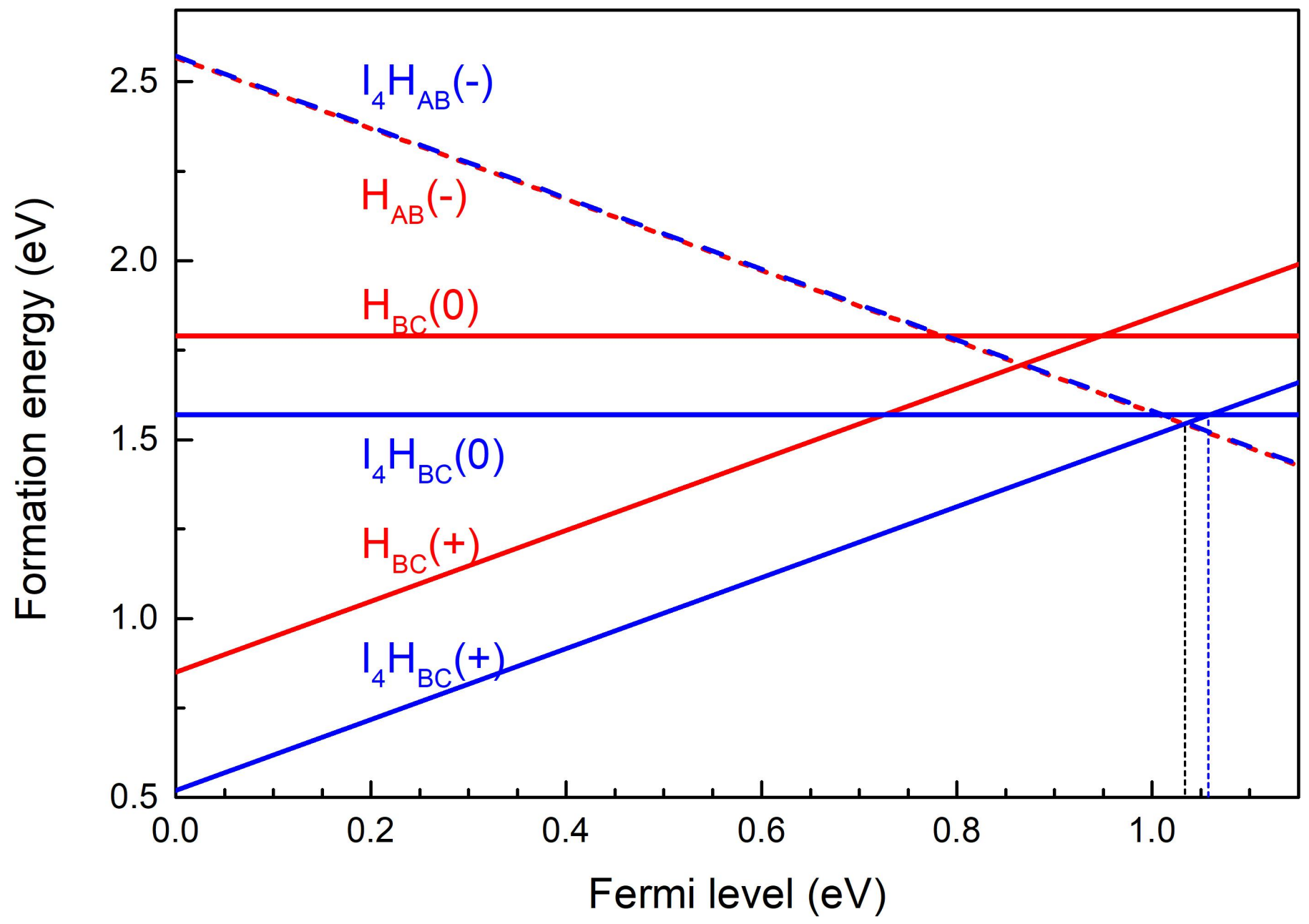

Fig. 2. Formation energy of interstitial hydrogen (red) with respect to pristine $Si_{512}$, and that of the $I_4H$ complex (blue) with respect to $Si_{512}$:$I_4$. The free $H_2$ molecule at 0K was taken as the hydrogen reservoir. The difference of the formation energies of $H_{AB}$(−) next to or far from $I_4$ is within the error bar of the calculations.

Fig. 2 compares formation energies of the most stable interstitial hydrogen defects in pristine Si and in the presence of $I_4$. As mentioned above, interstitial hydrogen in pristine silicon is a so-called negative-U defect, i.e., in equilibrium it can only be stable in the (+) and (–) charge states, occupying BC or AB positions, respectively.[61,70] In the presence of $I_4$, hydrogen is still bistable, preferring the $I_4H_{BC}$ configuration in the positive, and a $H_{AB}$ configuration outside the complex in the negative charge state. The negative-U behavior also prevails. Bistablility has been observed for several STDs, and interpreted as electron exchange with or hydrogen diffusion to a nearby vacancy,[71,72] or – as in the case of the SHD with a $C_iO_iH$ core – as local reconstruction of the defect configuration.[40] In case of the $I_4H$ complex the hydrogen atom is moving out of the BC position between self-interstitials to an AB position upon acquiring two electrons from some other defect, in a similar manner[45] as in the absence of $I_4$. According to calculations in the local density approximation (LDA), the energy barrier between $H_{BC}$(+) and $H_{AB}$(-) is ~0.3 eV.[73] A similar barrier can be expected from $I_4H_{BC}$ too. Note though that LDA usually underestimates energy barriers.)

While the calculated adiabatic CTLs are $E(+/0) = E_C - 0.21$ eV and $E(+/-) = E_V + 0.86$ eV for H in the pristine lattice (see SM), in the presence of $I_4$ we obtain $E(+/0) = E_C - 0.10$ eV and $E(+/-) = E_V + 1.03$ eV, respectively, i.e., much shallower than for H in the pristine lattice. $I_4H_{BC}(0)$ has a highly delocalized EMT-like donor state, so the size-convergent CTL is likely to be substantially shallower than 80 meV.[74] Our case studies prove, therefore, that the interaction of hydrogen with self-interstitial clusters produces shallow donors indeed when H inserts itself into the BC site of a stressed bond between the self-interstitials. At or below room temperature (RT), these shallow donors will be in a diamagnetic state and EPR could only detect them under illumination.

Kioyi and Umeda have found the MT2 center to have tetragonal symmetry, still assigning it tentatively to an $I_4H_2$ complex.[30] However, as mentioned earlier, two hydrogen atoms in conjunction with the $I_4$ complex can give rise to tetragonal symmetry only if both of them are on the main axis of $I_4$, but then they would form a diamagnetic $H_2$ molecule. Of course, $I_4H_{BC}$ does not have tetragonal symmetry either. However, it is conceivable that hydrogen hops between equivalent BC sites of $I_4$, causing EPR to detect a pseudo-tetragonal symmetry of the static center.

Therefore, we investigated such a migration paths of $H_{BC}(+)$. As shown in Fig. 3, the barrier between two $H_{BC}$ sites, as obtained by a three-image CI-NEB calculation with the PBE functional, is 0.49 eV. At first sight, this is too high for thermal migration at low temperature. Including the zero-point energy of the local vibrational modes (LVMs) calculated at the minimum and at the saddle point (PBE, H and its first and second neighbours are allowed to vibrate) reduces the barrier only slightly to 0.47 eV, because only the mode along the reaction coordinate ($\hbar\omega_{rc} \approx 32$ meV) is lost at the saddle point. A one-dimensional model of the periodic potential (see SM, Section S6) gives a thermal hopping rate below $10^{-30}$ $s^{-1}$ at cryogenic temperatures and a coherent ground-state tunnelling frequency ~ $10^{-15}$ Hz, so in the dark H is localized at one BC site. Under above-gap illumination, however, non-radiative capture and recombination at the defect release ≈ 1.1 eV locally, more than twice the barrier, and H can hop by the recombination-enhanced mechanism[75,76] with an estimated rate of $10^2$–$10^8$ $s^{-1}$, proportional to the excitation intensity. Since the calculated $^{29}$Si hyperfine tensors for the four equivalent BC sites differ by less than 0.3 MHz (Table S4), the deviation of the static $I_4H_{BC}(0)$ from tetragonal symmetry is expected to be below the EPR linewidth, and photo-assisted hopping averages it further, consistent with the tetragonal symmetry observed for MT2.

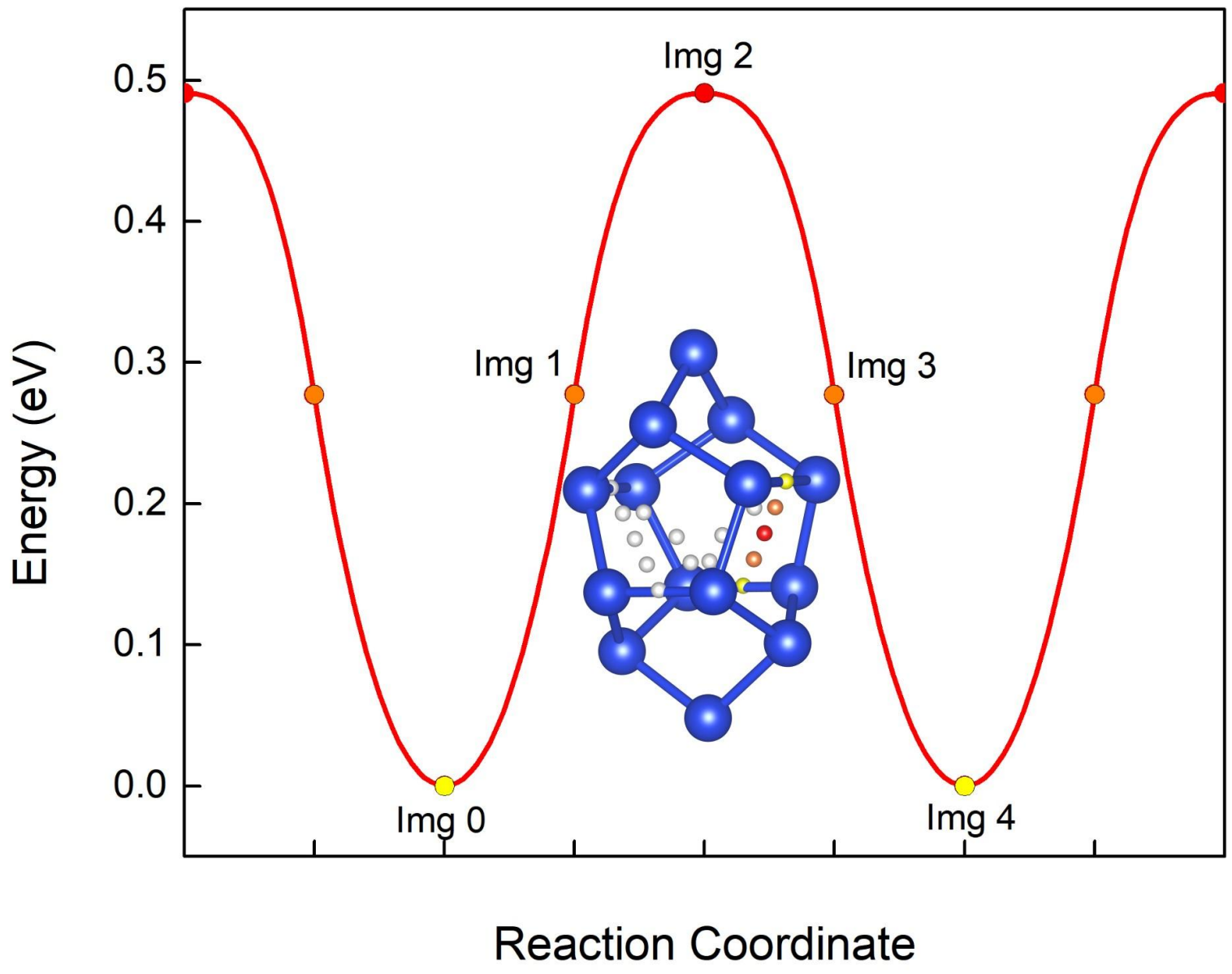

Fig. 3 Potential energy surface for the hopping of the $H_{BC}$ between equivalent BC sites. The curve is extended to show the periodic potential wells. The dots represent the NEB images computed from the three-image CI-NEB method. The inset shows the superimposed atomic structure illustrating all possible NEB paths between adjacent BC sites, where one path is colored to indicate the specific image, Img 2 (red dot) corresponds to the saddle point.

The hyperfine tensor measured for the MT2 center resembled qualitatively, but deviated quantitatively from the $B_3$ center, assumedly because of the presence of hydrogen, even though hf due to the hydrogen nuclear spin could not be detected.[30] We have calculated the hf tensor for all four equivalent BC positions of the $I_4H_{BC}$ defects and averaged them. As shown in Table I, our calculation gives a reasonable agreement with the experimental values on the MT2 center in the *a* shell and *b* shell of $^{29}$Si nuclei, but no detectable hyperfine coupling in the *c* shell. We note, however, that the measured shell-*a* tensor is more strongly anisotropic than the calculated tensor; this difference in anisotropy is plausibly due to the dipolar hyperfine contribution being sensitive to the delocalized EMT tail, which is truncated in the finite supercell. (We assume, that the assignment of 8 + 4 atoms to the second silicon shell in Ref.[30] might not have been warranted.) Even more importantly, our values calculated for hydrogen are also below the detection level, in agreement with the finding on MT2. Based on the calculated hf tensor, the MT2 center observed by Umeda and Kioyi can be specifically assigned to the $I_4H_{BC}$ complex.

**Table I.** Calculated hyperfine tensors of $I_4H_{BC}$, in comparison with experiment. All values in MHz. The silicon atoms in the core of the defect are classified, from top to bottom of Fig.1 as *a*, *b*, *c*, *c*, *b*, and *a*.

| | $I_4H_{BC}(0)$ calculated | | MT2[30] measured | |
|---|---|---|---|---|
| $^{29}$Si | $A_\parallel$ | $A_\perp$ | $\lvert A_\parallel\rvert$ | $\lvert A_\perp\rvert$ |
| *a* | –19.6 | –15.9 | 26.3 | 11.2 |
| *b* | –5.1 | –3.6 | 9.0 | 5.6 |
| *c* | –1.4 | –0.2 | -- | -- |
| H | –0.4 | –0.2 | -- | -- |

$I_4H_{BC}(0)$ also has a doubly occupied state localized on $I_4$, at 0.08 eV below the VBM. Since the shallow EMT donor state created by $H_{BC}$ is mostly ionized (giving rise to the observed free carriers), we may assume that excitation by light can also promote an electron from the $I_4$ state into the EMT state, and then recombination could give rise to PL, as depicted schematically in Fig. 4. By optimizing the geometry of the excited state of $I_4H_{BC}(+)$ using cDFT, we have calculated the ZPL energy to be 1128 meV after correction (see Table S5). Note that the ZPL can be written as $E_{\mathrm{ZPL}} = [E(+/0) - E_V] - E_b^h$, where $E_b^h$ is the binding energy of the hole to the neutral donor. Because the CTL and ZPL were not computed on equal footing and the relevant corrections are of the same order as the apparent $\mathrm{E}_{\mathrm{b}}^{\mathrm{h}}$, we conclude that the hole is at most weakly bound, as is also implied by the experimental data, where $E_g - h\nu$ of the $X_i$-type lines (20–46 meV) is comparable to the ionization energies of SHDs (34–53 meV).[39] This ZPL is close to energy range of the experimentally observed $X_i$ lines[29,31] and coincides with one of the peaks in the PL spectrum reported in Ref. [29] which anneals out in the second stage of the heat treatment. Therefore, our results identify the $I_4H$ complex as the origin of one of the SHDs with higher thermal stability. Invoking the results obtained on the IH+2H center, the insertion of a hydrogen atom into a Si–Si bond of an $I_n$ complex might give rise to other SHDs in that group.

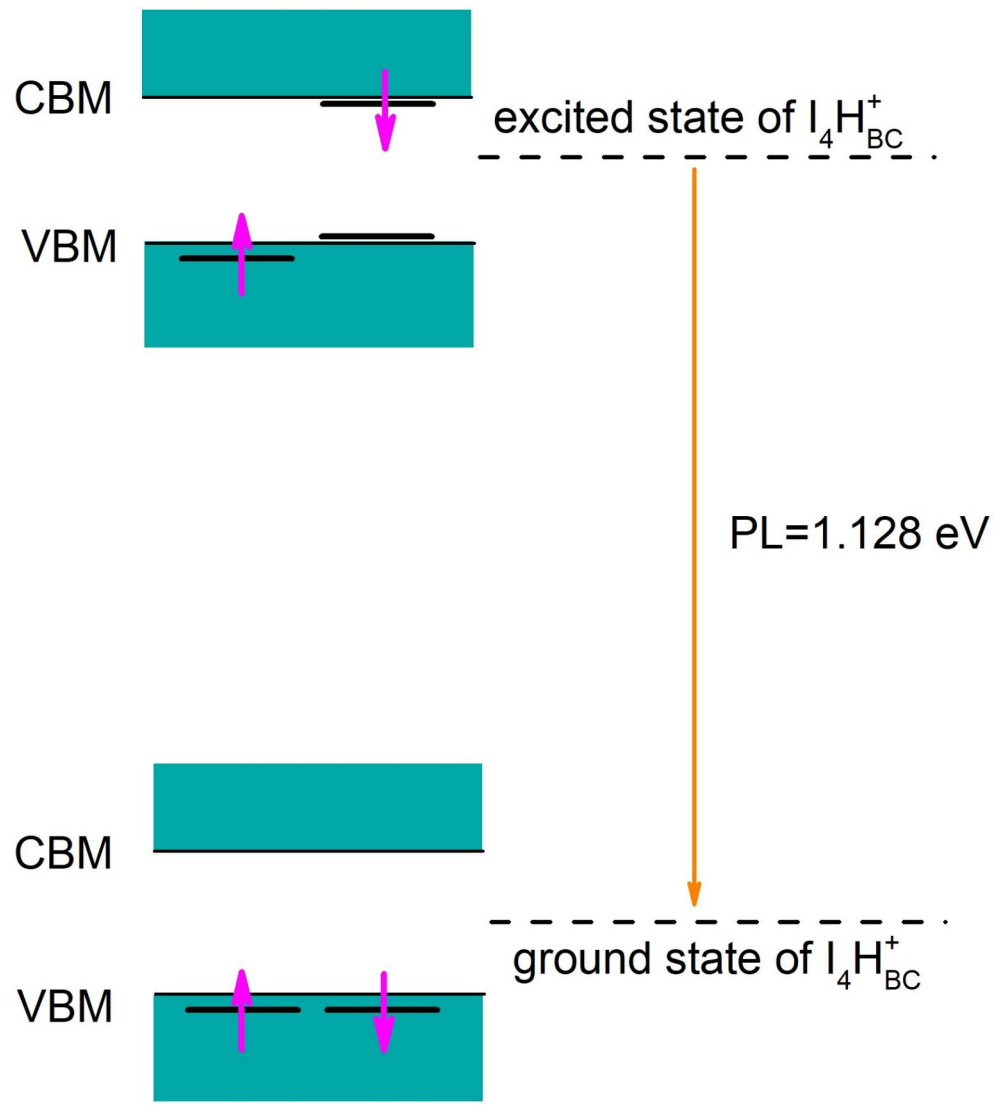

Fig. 4. Schematic energy level diagram showing PL from $I_4H_{BC}$.

## CONCLUSIONS

The EPR center MT2, experimentally observed in hydrogenated silicon, has been assigned to a shallow hydrogen related donor (SHD), tentatively assuming it be a negative-U, double donor $I_4H_2$ defect. Using HSE+$U_s$ calculations, we have shown here that hydrogen can insert itself as a bond centered (BC) interstitial into the compressed bonds of an $I_n$ complex, producing, indeed, a shallow, EMT-like single donor. Similarly to interstitial hydrogen in the pristine lattice, H can capture electrons from other defects and move out of the complex to an antibonding (AB) position in its negative charge state. Therefore, it can only be detected under illumination. Such a negative-U, bistable model for MT2 resolves the contradictions of the previous assignment, i.e., the number of observed free carriers per SHD, and the tetragonal symmetry, since we show that H can hop between the four equivalent BC sites of $I_4$ via photo-assisted mechanism. The calculated hf tensor for $I_4H_{BC}$ is also in good agreement with that of the MT2 center. The calculated PL of an $I_4H_{BC}$ complex is close to the experimentally observed $X_i$ lines and coincides with one shown to anneal out with MT2. Therefore, we identify MT2 with the shallow donor $I_4H_{BC}$ complex. Our work also establishes the $I_4H_{BC}$ complexes as the origin of one of the SHDs with high thermal stability (annealing out above 450 °C). Considering the results on the IH+2H complex, it appears likely that other SHDs, with the $X_i$ lines as their PL fingerprints, also arise when hydrogen attacks an $I_n$ cluster.

## ACKNOWLEDGMENT

M.H. gratefully acknowledges the support of the CSC scholarship (No. 202508330433), as well as the financial support from the Zhejiang Provincial Natural Science Foundation (ZCLY24A0401), Huzhou Natural Science Foundation (2024YZ24) and NSFC (Grant No. 61804154). The support from the Hefei Advanced Computing Center is also acknowledged. P.D. acknowledges the support of the NSFC grant No. 12250710132. B.H. acknowledges the National Key Research and Development of China (Grant No. 2022YFA1402400), NSFC (Grant No. W2511008, 12088101), and NSAF (Grant No. U2230402). A.G. acknowledges the high performance computational resources provided by KIFÜ (Governmental Agency for IT Development of Hungary) and the European Commission for the projects SPINUS (Grant No. 101135699), and QuSPARC(Grant No. 101186889).

## DATA AVAILABILITY

The authors declare that the main data supporting the findings of this study are available within the paper and its Supplementary files. Part of the source data is provided in this paper. The data that support the findings of this study are available from the corresponding author upon reasonable request.

SUPPLEMENTAL MATERIAL

# Origin and photoluminescence of a shallow hydrogen-related donor in silicon

Miaomiao Han,[1,2*] Peter Deák,[2,3] Bing Huang,[3] and Adam Gali[2,3,4*]

[1] School of Science, Huzhou Normal University, Huzhou, 313000 Zhejiang, China

[2] HUN-REN Wigner Research Center for Physics, Budapest, PoB 49, H-1525 Budapest, Hungary

[3] Beijing Computational Science Research Center, 100193 Beijing, China

[4] Department of Atomic Physics, Budapest University of Technology and Economics,H-1521 Budapest, Hungary

Corresponding authors: mmhan923@mail.ustc.edu.cn, gali.adam@wigner.hun-ren.hu

In defect calculations, the hybrid screened-exchange functional of Heyd, Scuseria, and Ernzerhof (HSE)[1,2,3] is often used to overcome the limitations of local (LDA) or semi-local (GDA) exchange in density functional theory (DFT). In most semiconductors, the two parameters of a HSE($\alpha,\mu$) functional ($\alpha$ for mixing non-local and semi-local exchange, and $\mu$ for the inverse screening length) can be tuned to reproduce the (0K) single-particle gap and to comply with the generalized Koopmans' theorem (gKT),[4] i.e., the ideal piecewise linearity of the exact xc energy function can be restored for the given material.[5] In fact, both parameters are related to electronic screening but the wave number dependence of the inverse dielectric function is still very crudely approximated in HSE, so that is why the optimal parameters will be materials-dependent.[6] The optimized parameter set can be determined based on the 0K single-particle band gap (known either from GW calculations or renormalizing observed optical absorption thresholds to 0K) and by checking either the deviation from linearity (i.e., the parabolicity) of the electron removal/addition energy, or by monitoring the gKT for an intrinsic defect. For silicon it has been shown, that HSE(0.25,0.20), often called HSE06, reproduces the gap accurately and satisfies the gKT within ~70 meV.[7] According to experience,[8,9,10] such parameters will work well for all defect states composed primarily of orbitals which are similar to those of the host material, e.g., for substitutional impurities with an *sp* valence shell in group IV or III-V semiconductors. Problems arise when, e.g., transition metal impurities are considered in these compounds, since their *d*-orbitals will change the screening locally. It has been shown that applying a Hubbard-*U* correction to the orbitals in question, i.e., a HSE($\alpha,\mu$)+$U_d$ functional, can account for that.[11] The value of *U* can be found either by enforcing the fulfillment of the gKT for the *d* state, or by comparing level positions to GW calculations.

It can be expected that screening around a hydrogen impurity (with a 1*s* valence shell) in silicon is not well described by the HSE06 parameters which are optimal for the host. Actually, in absence of intrinsic defects, the hydrogen atom prefers to sit into the bond between two Si atoms in its positive charge state, and behind a bond (antibonding, or AB site) when it is negative, and in both cases a three-center bond is formed, which is dominated by the orbitals of the two silicon atoms,[12] so the standard HSE parameters can give a fairly good account of the properties of interstitial hydrogen. However, when investigating a H atom trapped inside the cage of an $I_4$ self-interstitial complex, the possibility exists, that H resides in the center of the cage, in no covalent interaction with the Si atoms of the cage, and screening is determined more by the 1*s* state of hydrogen than by the $sp^3$ hybrids of the host. Therefore, it must be investigated, how the gKT is satisfied for a H atom fixed at the tetrahedral interstitial site of the pristine lattice.

A test on H fixed at the T-site of an unrelaxed $Si_{216}$ supercell has been conducted. H at the T site is expected to be a shallow acceptor, which can be nearly ideally described by effective mass theory (EMT). A pure HSE06 calculation for H fixed at the T-site of the unrelaxed silicon lattice gives results in stark contrast to this expectation. The gKT, i.e., the equality of the defect level position before and after ionization with the ionization energy (all referred to the VBM), is not satisfied either, as shown in Table S1.

**Table S1.** The gKT test for a H atom fixed at the interstitial T-site of an unrelaxed 216-atom Si supercell. Column 2 shows the Kohn-Sham level of the hole in the neutral charge state, while Column 4 shows it after occupation. Column 3 contains the ionization energy (as the difference of the self-consistent total energies of the two charge states). All values are given in [meV], with respect to the VBM. Total and one-electron energies of the negative state have been corrected with the SCPC (self-consistent potential correction) method.[13]

| U | LUMO(N) | E(N+1) – E(N) | HOMO(N+1) |
|---|---|---|---|
| 0.0 | 507 | 376 | 98 |
| -5.0 | 246 | 217 | 39 |
| -7.0 | 158 | 153 | 16 |
| -8.5 | 45 | 113 | 40 |
| -11.5 | 45 | 113 | 40 |

Application of $U_s = -8.5$ eV for hydrogen in the HSE calculation fixes both issues. The acceptor state is then close to the VBM, and the gKT is satisfied within 70 meV. Note that the H 1*s* level becomes doubly occupied and below the VBM at this $U$ value, giving rise to a VBM-derived EMT acceptor state, as expected. Therefore, upon further increase in the absolute value of $U$, we are not testing the gKT on the H atom but on the lattice. The 70 meV error in satisfying the gKT is roughly the same as HSE06 has for a series of substitutional EMT dopants.[11]

We have tested the HSE(0.25,0.2) + $U_s = -8.5$ eV functional in the experimentally well characterized equilibrium states of the hydrogen interstitial in the silicon lattice, applying a 512-atom supercell in the $\Gamma$-point approximation. The charge transition levels have been measured at room temperature (RT) to be $E(+/0) = E_{CBM} - 0.18$ eV (at the bond-center, or BC site)[14] and $E(+/-) = E_{VBM} + 0.76$ eV (between the BC and AB sites).[15,16] For comparison with our calculation, relevant for 0K, in Table S2 we have upscaled these values with the ratio of the low-temperature and RT gaps of silicon, 1.17/1.12.

**Table S2.** Charge transition levels of the hydrogen interstitial in pristine Si, in equilibrium at the BC site for the positive and neutral, and at the AB site in the negative charge state. All values in [eV].

| U | (+/–) | (+/0) |
|---|---|---|
| 0.0 | VBM + 0.67 | CBM – 0.25 |
| –7.0 | VBM + 0.85 | CBM – 0.22 |
| –8.5 | VBM + 0.86 | CBM – 0.21 |
| –11.0 | VBM + 0.88 | CBM – 0.25 |
| Experimental* | VBM + 0.79 | CBM – 0.19 |

* The RT experimental values of Refs.[14,15,16] have been upscaled with the ratio of the low-temperature and RT gaps of silicon.

It is clear that increasing the absolute value of $U_s$ beyond 8.5 eV increases the deviation from experiments. Decreasing it, however, increases the error in satisfying the gKT (see Table S1). Note that – unlike in the case of H at the T-site, the states relevant to these charge transitions are mostly dominated by silicon orbitals, so the application of the $U$ term does not have a very profound influence, as expected, but it does improve the agreement with experiment. Therefore, in the paper, we used $U_s = -8.5$ eV considering that it gives the smallest deviation in the gKT test, the best values for interstitial H, and also it predicts $H_T$ to be an EMT acceptor (as it should be).

To demonstrate that hydrogen at the BC site in one of the compressed Si–Si bonds of a self-interstitial complex ($I_n$) will be shallower than $H_{BC}$ in the pristine lattice, we adopt the single self-interstitial $I_{[001]}H_2$ configuration and insert one hydrogen between the two Si atoms (see Fig. S1) . We find that a shallow donor level arises, 60 meV below the CBM. The calculated hyperfine tensor of $I_{[001]}H_2H_{BC}^+$ is shown in Table S3.

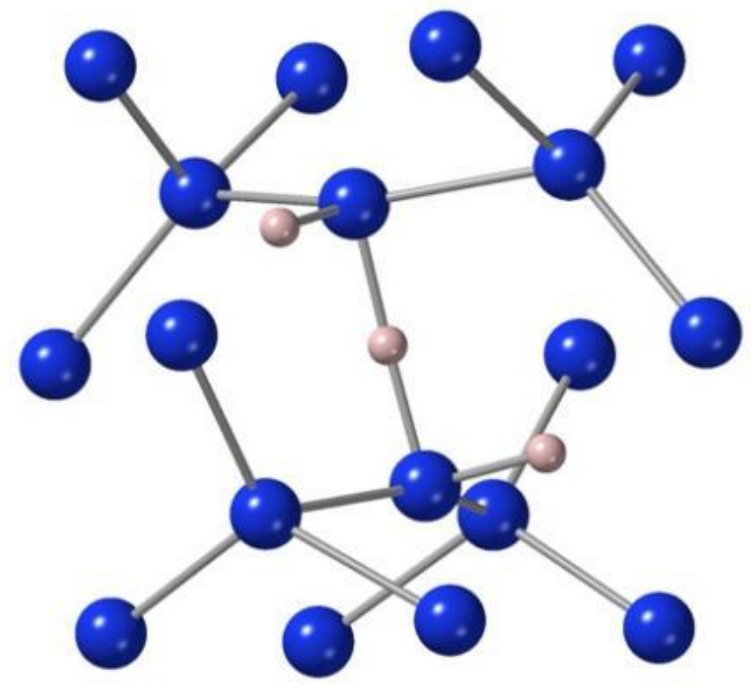

**Fig. S1.** The $I_{[001]}H_2H_{BC}$ defect.

**Table S3**. Calculated hyperfine tensor of $I_{[001]}H_2H_{BC}$. All values in MHz.

| ion | $A_{xx}$ | $A_{yy}$ | $A_{zz}$ |
|---|---|---|---|
| $H_1$ | –0.8 | –0.2 | –0.8 |
| $H_2$ | –0.8 | –0.2 | –0.8 |
| $H_{BC}$ | 16.9 | 16.9 | 18.5 |
| $Si_1$ | –43.0 | –40.4 | –48.1 |
| $Si_2$ | –43.0 | –40.4 | –48.1 |

**Table S4**. Calculated hyperfine tensor of $I_4H_{BC}^0$ with H occupying each of the four symmetry-equivalent BC sites. All values in MHz.

| | $I_4H_{BC}$(0) -Conf.1 | | $I_4H_{BC}$(0) -Conf.2 | | $I_4H_{BC}$(0) -Conf.3 | | $I_4H_{BC}$(0) -Conf.4 | |
|---|---|---|---|---|---|---|---|---|
| $^{29}$Si | $A_{\parallel}$ | $A_{\perp}$ | $A_{\parallel}$ | $A_{\perp}$ | $A_{\parallel}$ | $A_{\perp}$ | $A_{\parallel}$ | $A_{\perp}$ |
| *a* | -19.77 | -16.03 | -19.57 | -15.88 | -19.64 | -15.94 | -19.53 | -15.86 |
| *b* | -5.16 | -3.68 | -5.10 | -3.61 | -5.12 | -3.64 | -5.09 | -3.60 |
| *c* | -1.38 | -0.20 | -1.54 | -0.23 | -1.39 | -0.23 | -1.41 | -0.25 |
| H | -0.38 | -0.23 | -0.38 | -0.23 | -0.38 | -0.23 | -0.38 | -0.23 |

Table S5 shows the raw PL energy and the corresponding corrections (as described in the computational methods section) for the $I_4H$ complex,.

**Table S5**. Calculated PL energy and the corresponding corrections, all in meV.

| Raw value | Charge correction | Spin correction | Band -filling | Singularity |
|---|---|---|---|---|
| 1115 | 12.8 | 9.3 | –47.3 | 38 |

**S6 Hopping rate of H between equivalent BC sites.** The minimum-energy path of Fig. 3 is approximated by $V(x) = V_0/2\,[1 - \cos(2\pi x/L)]$, where $L$ is the distance between adjacent BC sites along the path. The harmonic frequency at the minimum is $\hbar\omega_{rc} = 2\sqrt{V_0 E_R}$ with $E_R = \hbar^2\pi^2/2mL^2$; hence the Schrödinger equation reduces to the Mathieu equation with $q = V_0/4E_R = (V_0/\hbar\omega_{rc})^2$, and the spectrum depends only on $V_0/\hbar\omega_{rc}$. The eigenvalues were obtained in a plane-wave basis for Bloch wave vectors $k = 0$ and $k = \pi/L$; the width $W_n$ of the n-th band gives the coherent tunnelling frequency $W_n/h$ of level *n* in a ring of equivalent sites (an upper bound to the incoherent hopping rate). The thermal rate is the Boltzmann-weighted sum of these frequencies for levels below the barrier top, and $k_BT/h$ above it. With $V_0 = 0.49$ eV and $\hbar\omega_{rc}$ = 32 meV, $q = 235$, and fifteen levels lie below the barrier top; the tunnelling frequencies are $3{\times}10^{-15}$, $4{\times}10^{-13}$, $8{\times}10^{-11}$, $2{\times}10^{-8}$, $3{\times}10^{-6}$, $5{\times}10^{-4}$, $8{\times}10^{-2}$, $1{\times}10^{1}$, $2{\times}10^{3}$, $3{\times}10^{5}$, $4{\times}10^{7}$, $4{\times}10^{9}$, $1{\times}10^{11}$, $2{\times}10^{12}$ and $5 \times 10^{12}$ Hz for n = 0 … 14. The crossover temperature is $T_c = \hbar\omega_{rc}/2\pi k_B$ = 58 K.

**Table S6.** Sensitivity of the model to the barrier height and reaction-mode frequency (rates in s$^{-1}$).

| $V_0$ (eV) | ħω$_{rc}$ (meV) | ZPE-corr. barrier (eV) | ground-state tunnelling freq. (Hz) | $k_{th}$(150 K) | $k_{th}$(300 K) |
|---|---|---|---|---|---|
| 0.40 | 32 | 0.38 | $< 10^{-20}$ | $4\times10^{-1}$ | $2\times10^{6}$ |
| 0.49 | 20 | 0.48 | $< 10^{-30}$ | $2\times10^{-4}$ | $3\times10^{4}$ |
| 0.49 | 32 | 0.47 | $3\times10^{-15}$ | $4\times10^{-4}$ | $5\times10^{4}$ |
| 0.49 | 40 | 0.47 | $1\times10^{-9}$ | $5\times10^{-4}$ | $6\times10^{5}$ |
| 0.49 | 60 | 0.46 | $1\times10^{0}$ | $1\times10^{-3}$ | $1\times10^{5}$ |
| 0.60 | 32 | 0.58 | $< 10^{-20}$ | $7\times10^{-8}$ | $7\times10^{2}$ |

($k_{th}$ = harmonic TST with ZPE; the tunnelling column is the coherent upper bound.)

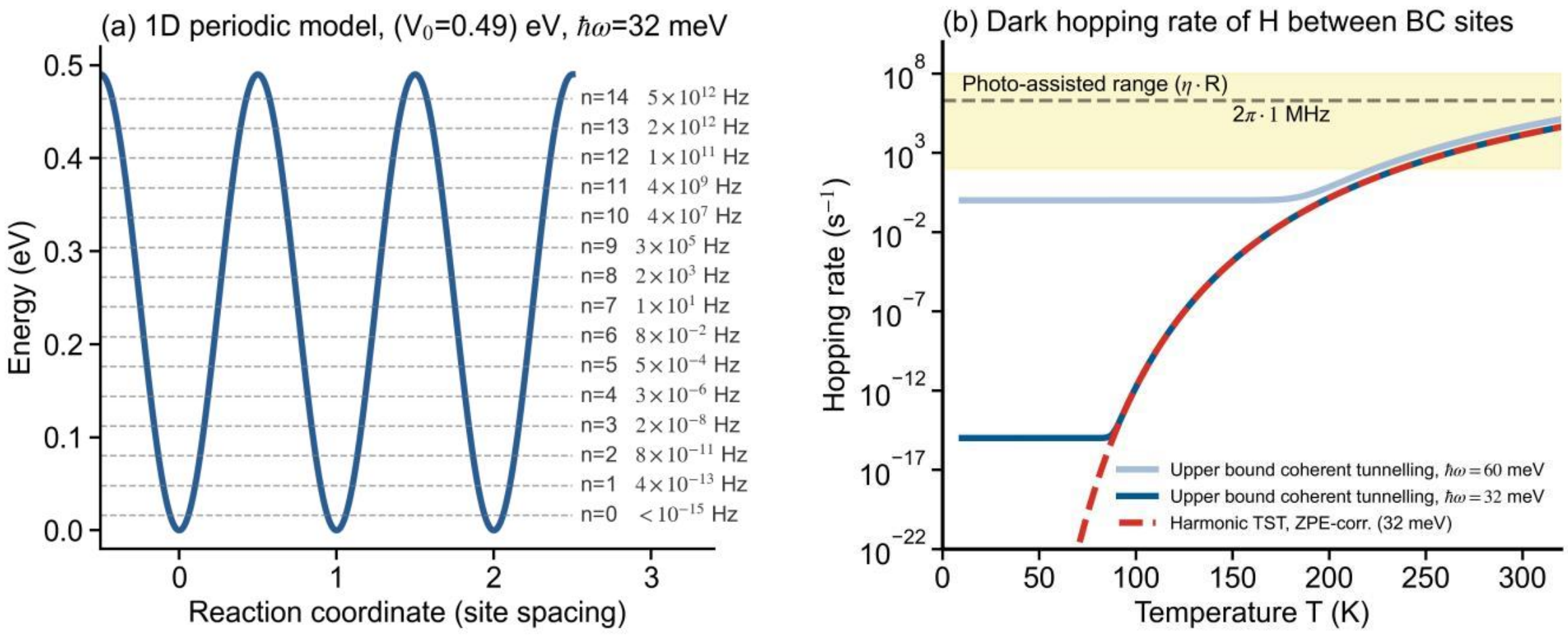

**Fig. S2.** (a) Periodic model potential with the vibrational levels of the reaction mode and their tunnelling frequencies. (b) Dark hopping rate vs temperature: harmonic transition-state theory with ZPE (dashed) and the upper bound including coherent tunnelling for two reaction-mode frequencies. The shaded band is the estimated photo-assisted rate under typical EPR illumination.